\documentclass{aa}
\usepackage{aas_macros}

\usepackage[utf8]{inputenc}
\usepackage{graphicx}
\usepackage[export]{adjustbox}
\usepackage{stackengine}
\usepackage{epstopdf}
\usepackage{url}
\usepackage{natbib}
\bibpunct{(}{)}{;}{a}{}{,}
\usepackage{xcolor}
\usepackage{multirow}
\usepackage{textcomp,gensymb}
\usepackage{float}
\usepackage{dblfloatfix}
\usepackage[percent]{overpic}
\usepackage{array}
\usepackage{booktabs}
\usepackage[switch]{lineno}
\usepackage[hidelinks,colorlinks=true,linkcolor=blue,citecolor=blue,urlcolor=blue]{hyperref}

\usepackage{subcaption}
\usepackage{siunitx}
\newcommand{\degrees}{\ensuremath{^\circ}}
\newcommand{\hours}{\ensuremath{^\mathrm{h}}}
\newcommand{\minutes}{\ensuremath{^\mathrm{m}}}
\newcommand{\seconds}{\ensuremath{^\mathrm{s}}}
\newcommand{\PSRB}{PSR B1853+01}
\usepackage{threeparttable}
\begin{document}
    \title{The nebula around pulsar B1853+01 in X-rays}
    \titlerunning {the PWN around B1853+01}
    \author{Xiying Zhang\inst{1}\fnmsep\thanks{e-mail: xzhang@icc.ub.edu}, Pol Bordas\inst{1}, Samar Safi-Harb\inst{2} \and Kazushi Iwasawa\inst{1,3}}
    \institute{Departament de Física Quàntica i Astrofísica, Institut de Ciències del Cosmos, Universitat de Barcelona, IEEC-UB, Martí i Franquès 1, 08028, Barcelona, Spain \and Department of Physics and Astronomy, University of Manitoba, Winnipeg, MB, R3T 2N2, Canada \and ICREA, Pg Lluís Companys 23, 08010 Barcelona, Spain}
    \date{Received --- ; accepted ---}

\abstract{
We report on the results of a comprehensive analysis of X-ray observations with \textit{Chandra}, \textit{XMM-Newton} and \textit{NuSTAR} of the pulsar wind nebula (PWN) associated with PSR B1853+01, located inside the W44 supernova remnant (SNR). Previous X-ray observations unveiled the presence of a fast-moving pulsar, PSR B1853+01, at the southern edge of the W44 thermal X-ray emission region, as well as an elongated tail structure trailing the pulsar. Our analysis reveals, in addition, an ``outflow'' feature ahead of the pulsar extending for about 1\arcmin~($\sim$1.0 pc at a distance of 3.2 kpc). 
At larger scales, the entire PWN seems to be surrounded by a faint, diffuse X-ray emission structure. The southern part of this structure displays the same unusual morphology as the ``outflow'' feature and extends along $\sim$6\arcmin~($\sim$5 pc) in the direction of the pulsar proper motion. In this report, a spatially-resolved spectral analysis for different extended regions around PSR B1853+01 is carried out. For an updated value of the column density of $0.65_{-0.42}^{+0.46} \times 10^{22} ~\textrm{cm}^{-2}$, a power-law fit to the ``outflow'' region yields a spectral index $\Gamma \approx 1.24_{-0.24}^{+0.23}$, which is significantly harder than that of the pulsar ($\Gamma \approx 1.87_{-0.43}^{+0.48}$) and the pulsar tail ($\Gamma \approx 2.01_{-0.38}^{+0.39}$). We argue that both the ``outflow'' structure and the surrounding halo-like X-ray emission might be produced by high-energy particles escaping the PWN around PSR B1853+01, a scenario recently suggested also for other bow-shock PWNe with jet-like structures and/or TeV halos.}
\maketitle

\section{Introduction}
Rotation-powered pulsars spin down and lose a large fraction of their energy through relativistic magnetized winds. Launched from the pulsar magnetosphere, pulsar winds with supersonic bulk velocities are thought to be composed, mostly, of electron-positron pairs and electromagnetic field \citep{Goldreich1969ApJ,Sturrock1971ApJ,Ruderman1975ApJ,Kirk2009}. The interaction between the pulsar wind and the ambient medium, being the supernova ejecta if the pulsar still resides inside the SNR, or the interstellar medium (ISM) for a runaway pulsar, produces a termination shock where the pulsar wind is thermalized. Downstream the shock, particles accelerated to relativistic energies can emit broadband synchrotron and inverse Compton (IC) radiation forming what we observe as a pulsar wind nebula (PWN) \citep{Rees_and_Gunn_1974,1984KC_Confinement,1984KC_MHDCrab}. In X-rays, where synchrotron radiation dominates, footprints of ``fresh'' electrons and positrons with energies of $\sim$10 to 100 TeV can be captured using high resolution satellite-based X-ray observatories. So far $\sim$80 PWNe systems have been discovered in the past two decades in this energy band \citep{Kargaltsev2008}.

The morphology of PWNe strongly depends on the proper motion of the pulsar. Deep observations combined with numerical simulations of subsonically moving pulsars (e.g. the Crab pulsar) show that outside the light cylinder, the energy flux of pulsar winds tends to concentrate preferentially close to the equatorial plane, whereas downstream the termination shock, the shocked equatorial wind is partly diverted towards the pulsar poles, forming the so-called torus-jet structure \citep{Bogovalov2002MNRAS, Lyubarsky2002MNRAS}. In the case of pulsars moving faster than the local sound speed, the PWN morphology is heavily affected by the pulsar's proper motion, featuring a bow shock head and cometary tail structure with the pulsar located close to the head. Based on their morphology, supersonically moving PWNe are dubbed bow-shock PWNe (BSPWNe). 

\begin{figure*}[t!]
     \begin{subfigure}[h]{0.49\linewidth}\includegraphics[width=\linewidth]{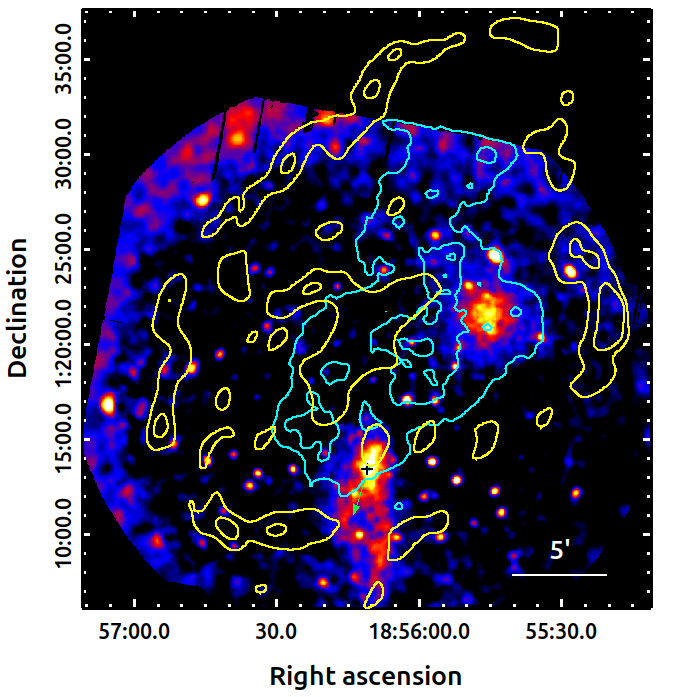}
    \end{subfigure}
     \begin{subfigure}[h]{0.49\linewidth}\includegraphics[width=\linewidth]{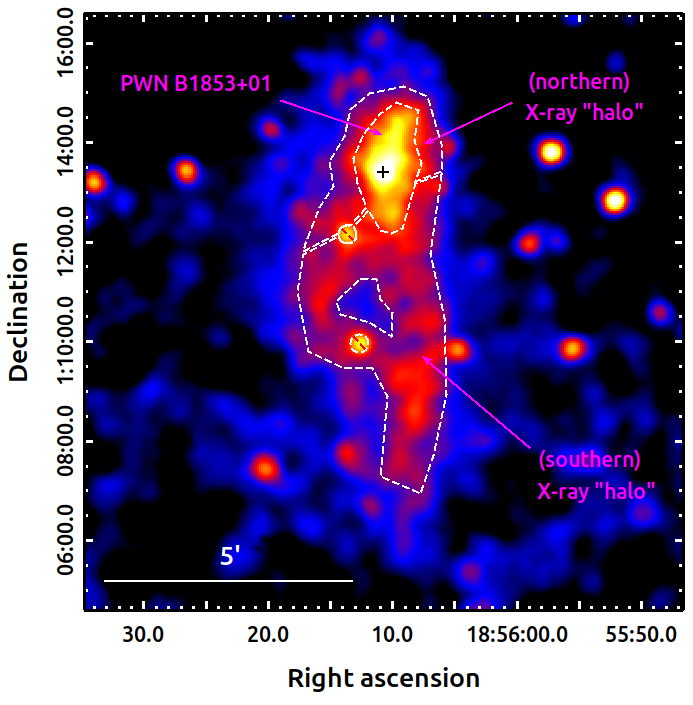}
    \end{subfigure}
     \caption{\textit{XMM-Newton} exposure-corrected mosaic images of SNR W44 region in hard (4.0-8.0 keV) X-ray bands. \textbf{Left}: In the FoV of W44 with the SNR radio contours in yellow (radio fits file obtained from SNRCat at {\url{http://snrcat.physics.umanitoba.ca/SNRrecord.php?id=G034.7m00.4}}) and the soft (0.5-4.0 keV) X-ray contour of W44 in cyan. The position of PSR B1853+01 is marked by a black cross with a green arrow indicating the direction of the pulsar's proper motion inferred from the pulsar's position relative to the SNR center and the axis of symmetry of the PWN.
     \textbf{Right}: 
     Extended X-ray emission coinciding with the position of PSR B1853+01 inside of W44. The position of PSR B1853+01 is marked by a black cross. The inner part of the dashed white lines corresponds to the PWN surrounding PSR B1853+01 while the faint extended emission surrounding the PWN is a possible X-ray "halo". Note that the edge of the thermal emission from SNR cut the halo into two parts. }
    \label{fig:W44}
\end{figure*}

Recent observations performed with \textit{Chandra} \citep{Kargaltsev2017} suggest that it is more common than previously expected that BSPWNe tend to be accompanied with peculiar misaligned ``outflows'' or filamentary features. The extension and orientation of these features is difficult to accommodate in a ballistic-jet scenario, prompting for alternative mechanisms accounting for their nature based on the escape of high-energy particles accelerated at the PWN bow shock (\citealp{Bandiera2008}). Meanwhile, extended emission at very high energy (VHEs, $E > 100$~GeV) gamma-rays has been recently unveiled by Cherenkov detectors around middle-aged pulsars. The so-called TeV halos \citep{Abeysekara2017,YiZhang2021PhRvL} are now identified as a new gamma-ray source class \citep{Linden2017PhRvD}, as anticipated in \cite{Aharonian1995}. In this scenario, the gamma-ray emission is interpreted as IC on CMB photons by electrons/positrons that escape the PWNe and diffuse into the ISM. Yet, many properties of these TeV halos remain unknown, for which new insights provided by the detection of extended X-ray emission structures around PWNe could be crucial. 

W44 (G34.7-0.4) is a composite, mixed-morphology SNR formed about $\sim$10 kyr ago \citep{Smith1985MNRAS} and located at a kinematic distance of $\sim$3.0 kpc, based on OH and HI absorption measurements \citep{Casewll1975A&A, Claussen1997ApJ, Ranasinghe2018AJ}. At radio wavelengths, W44 presents a distorted elliptical shell-like shape of about 25\arcmin $\times$35\arcmin~with the bulk of the radio emission concentrated in knots and filamentary structures \citep{Clark1975AuJPA,Jones1993MNRAS,Frail1996ApJ}. Inside the radio shell, due to the presence of overionized recombining plasma \citep{Shelton2004ApJ,Uchida2012PASJ,Okon2020ApJ}, the remnant is filled with thermal X-ray emission which displays an elongated shape in the NW-SE direction \citep{Watson1983IAUS,Rho1994ApJ,Harrus1996}, as shown in the left panel of Fig. \ref{fig:W44}. 
Hard X-ray imaging of W44 \citep{Harrus1996,Uchida2012PASJ,Okon2020ApJ}, however, showed two distinct components in the field of the SNR: a diffuse emission towards the west possibly associated to a galaxy cluster behind the Galactic plane \citep{Uchida2012PASJ,Nobukawa2018ApJ}, and an extended structure towards the south close to the position of the PWN associated with PSR B1853+01 (see Fig. \ref{fig:W44}). In addition, two extended GeV gamma-ray sources were found at the two opposite edges of the SNR major axis \citep{Uchiyama2012ApJ,Peron2020ApJ}, where the shell of W44 might be interacting with nearby molecular clouds \citep{Wootten1977ApJ,Seta2004AJ}. At VHEs, the MAGIC and VERITAS Cherenkov telescopes have recently reported upper limits to the TeV emission from the SNR \citep{MAGICW442025A&A...693A.255A,VERITASW442025ApJ...983...73A}.

The 267 ms pulsar PSR B1853+01/PSR J1856+0113 \citep{Wolszczan1991ApJ} is located south of W44, well within its radio shell and at the southern edge of the thermal X-ray emission region. PSR B1853+01 slows down at a rate $\dot{P} \sim 208 \times 10^ {-15} \textrm{ s s}^{-1}$, which leads to a characteristic age $\tau_c \sim 20$~kyr and a spin-down luminosity $\dot{E}\sim 4.3\times 10^{35} \textrm{ erg s}^{-1} $. Based on its dispersion measure, the distance to the pulsar is estimated to be $\sim$3.2 kpc \citep{Wolszczan1991ApJ}. Further pulsar-related parameters are listed in Table \ref{tab:pulsar info}. The similarities in both age and distance for PSR B1853+01 and SNR W44 strongly suggest an association between them. 
\begin{table}
    \centering
    \caption{Parameters of PSR B1853+01}
    \label{tab:pulsar info}
    \begin{adjustbox}{width=1.0\linewidth}
    \renewcommand{\arraystretch}{1.3}
    \begin{tabular}{c c}
    \toprule
    \toprule
         Parameters \\
         \midrule
         R.A. (J2000) &  18\hours56\minutes10\seconds89 $\pm$ 0\seconds.01\\
         Decl. (J2000) &  01\degrees13\arcmin20\arcsec.6  $\pm$ 0\arcsec.3\\
         P(s)& $0.26739884073$ $\pm$ 0.00000000003\\
         $\Dot{P} (\textrm{s s}^{-1})$& (208.408 $\pm$ 0.003)$\times 10 ^{-15}$ \\
         \midrule
         Derived Parameters \\
         \midrule
         Magnetic field & $7\times 10 ^{12} \textrm{ G}$\\
         Characteristic age & $2\times 10^{4} \textrm{ yr}$\\
         Spin-down luminosity& $4.3\times 10^{35} \textrm{ ergs s}^{-1} $\\
         Distance &  3 kpc\\ 
         Velocity$^a$ &  375 km/s\\
    \bottomrule
    \end{tabular}
    \end{adjustbox}
    \tablefoot{Table adapted from \citet{Wolszczan1991ApJ}. \tablefoottext{a}{Reference: \citet{Frail1996ApJ}}}
\end{table}

An extended nebula around PSR B1853+01 was first revealed by \citet{Jones1993MNRAS} in the radio band. \citet{Frail1996ApJ} confirmed the PWN nature of the extended radio emission based on 1) the detection of the pulsar at the apex of a cometary-shaped emission with an extension of about 2.5\arcmin~(or about 2.3 pc at $d =$ 3.2 kpc) and 2) a relatively flat spectral index $\alpha=-0.12\pm0.04$ (where $S_{\nu}={\nu}^{\alpha}$) for the nebula, compared to that for the rest of the SNR ($\alpha=-0.33$). \citet{Harrus1996} reported the detection of the PWN using ASCA observations (4.0-9.5 keV) where W44 is invisible and a source coincident with the position of the pulsar is evident. High-resolution X-ray observations with $Chandra$ further revealed the cometary tail structure of the PWN trailing the pulsar \citep{Petre2002ApJ}.

In this paper we present a detailed X-ray study of PSR B1853+01 and its surrounding nebula based on spatially resolved spectral analysis of this system using archival data from \textit{Chandra}, \textit{XMM-Newton} and \textit{NuSTAR}. The structure of this paper is as follows: in Section 2 we describe the observations used in this work and the data reduction processes. In Section 3 we present our imaging and spectral analysis results, including images obtained from different instruments featuring structures at different length-scales, together with improved constraints on their spectral properties. In Section 4 we discuss the obtained results, focusing on the nature of the extended emission around and beyond the PWN of PSR B1853+01, and compare our findings to other similar BSPWNe. Finally in Section 5 we summarize our results and conclusions.

\section{Observations and Data Reduction}
\begin{table*}
\centering
\begin{adjustbox}{ width=1.0\textwidth}
\begin{threeparttable}
\caption{Observation log of the \textit{XMM-Newton},\textit{Chandra}, and \textit{NuSTAR} data used for analysis.}
\renewcommand{\arraystretch}{1.5}
    \begin{tabular}{ lcccccccc} 
    \toprule
    \toprule
    & Obs. ID& Instrument & Mode$^a$ &Start Date  & Pointing & Effective Exposure & PI & Ref.\tnote{e}\\
    \midrule
    \multirow{3}{*}{\textit{Chandra}}& 1958 &ACIS-S& FAINT&2000-10-31 &  W44 center\tnote{b} & 44.79 ks&Shelton &[1] \\ 
    & 5548 &ACIS-S& VFAINT & 2005-06-25 & W44 PWN & 52.31 ks& Petre  & [2]\\
    & 6312 &ACIS-S& VFAINT & 2005-06-23 & W44 PWN & 39.58 ks& Petre  & [2]\\
    \midrule
    \multirow{4}{*}{\textit{XMM-Newton}} &0551060101 & MOS1/MOS2/PN&FF/FF/FF&2009-04-24 & W44 PWN & 56.3/64.0/31.6 ks& Petre & [2]\\
        & 0721630101 & MOS1/MOS2/PN&FF/FF/EFF&2013-10-18 & W44\tnote{c}& 107.9/108.9/87.4 ks& Smith & [2]\\
        & 0721630201 & MOS1/MOS2/PN&FF/FF/EFF&2013-10-19 & W44\tnote{c}& 94.1/94.0/74.0 ks& Smith & [2]\\
        & 0721630301 & MOS1/MOS2/PN&FF/FF/EFF&2013-10-23 & W44\tnote{c}& 93.6/94.5/66.3 ks& Smith & [2]\\
    \midrule
    \textit{NuSTAR} & 40401005002 & FPMA+FPMB &&2018-09-02 &South of W44\tnote{d} & 105.19/104.66 ks& Uchida & [2] \\
    \bottomrule
    \end{tabular}
  \tablefoot{ $\large^a$FF: Full Frame, EFF: Extended Full Frame ~$\large^b$off-axis angle $\sim7\arcmin$ ~$\large^c$off-axis angle $\sim4.5\arcmin$ ~$\large^d$off-axis angle $\sim2.2\arcmin$ ~$\large^e$Here we only list references related to the study of the PWN associated with PSR B1853+01. [1] \citealp{Petre2002ApJ}; [2] this work }
\end{threeparttable}
\end{adjustbox}

\end{table*}\label{tab:Observation_List}

\subsection{\textit{XMM-Newton}}
There are in total seven archival observations where PSR B1853+01 and its nebula are found in \textit{XMM-Newton}'s field of view (FOV). Our study is based on the analysis of four of these archival observations (Obs. ID=0551060101, 0721630101, 0721630201, and 0721630301), as the three remaining pointings accounted for very short exposures which could lead to inaccurate results due to flaring background. Details of the used observations are given in Table \ref{tab:Observation_List}. These observations were taken with the European Photon Imaging Camera (EPIC) onboard \textit{XMM-Newton}, composed of three cameras: MOS1, MOS2 and \textit{pn} \citep{Turner2001,Struder2001}. EPIC cameras allow for highly sensitive imaging observations over the telescope's FOV of 30\arcmin~in the energy range from 0.15 to 15 keV with moderate spectral ($E/\Delta E\sim$ 20-50) and angular ($\sim$6\arcsec~FWHM of the PSF\footnote{{\url{http://xmm-tools.cosmos.esa.int/external/xmm_user_support/documentation/uhb/}}}) resolution. We note that Obs. 0551060101, the only on-axis observations used in this paper, was performed on April 24th 2009, after the loss of MOS1 CCD6, whereas for Obs. 0721630101, 0721630201, and 0721630301, these observations were taken with an off-axis angle about 4.5\arcmin~on 18th, 19th and 23th of October 2013, respectively, when neither MOS1 CCD6 nor MOS1 CCD3 were available.

The analysis reported in this work has been performed using the \textit{XMM-Newton} Science Analysis System\footnote{{\url{https://xmm-tools.cosmos.esa.int/external/xmm_user_support/documentation/sas_usg/USG/}}} (SAS) version 20.0.0, on an Ubuntu 20.04 platform. We also made use of the \textit{XMM-Newton} Extended Source Analysis Software (XMM-ESAS) and related analysis procedures\footnote{{\url{https://heasarc.gsfc.nasa.gov/docs/xmm/xmmhp_xmmesas.html}}}, since we are mostly interested in the extended emission around and beyond the PWN of \PSRB.

To reduce the data retrieved from the \textit{XMM-Newton} Science Archive, \texttt{emchain} and \texttt{epchain} were firstly applied to both MOS and \textit{pn} observations to filter events based on good time intervals, malfunctioning pixels, and out-of-time events. The task \texttt{espfilt} was then applied to filter out cosmic soft proton events (for \textit{pn} we set \texttt{rangescale=15.0}).

For the imaging analysis, we generated the spectra for the entire region, the redistribution matrix files and the ancillary response files, using \texttt{mos-spectra} and \texttt{pn-spectra}, for MOS and \textit{pn} detectors, respectively. We estimated the non-X-ray background (NXB) with \texttt{mos-back} and \texttt{pn-back}. The \texttt{merge$\_$comp$\_$xmm} tool was then used to merge the individual components (counts, exposure, particle background) from the four observation data sets. We then obtained background subtracted, exposure corrected, and adaptively smoothed images using \texttt{adapt$\_$merge}. For the spectral analysis, we first used \texttt{especget} to extract spectra for individual pointngs. Then \texttt{epicspeccombine} and \texttt{grppha} were used to combine and group the spectra.

\subsection{\textit{Chandra}}

Three observations (ObsID. 1958, 5548, 6312) containing PSR B1853+01 and its nebula were found in \textit{Chandra}'s archive (see Table \ref{tab:Observation_List} for details). These observations were performed with the Advanced CCD Imaging Spectrometer (ACIS) \citep{Gamire2003}, in Timed Exposure (\texttt{TE}) mode,
without using any gratings. ACIS features two arrays of CCDs: ACIS-I, optimized for imaging wide fields, and ACIS-S, the one used in this study, which provides high-resolution on-axis imaging capabilities. ACIS-S consists of 6 chips (S0 - S5; S1 and S3 chips being back illuminated) and has a field of view (FOV) of 8.4\arcmin$\times$51.1\arcmin. ACIS-S operates in the energy range 0.2-10 keV, with an energy resolution $E/\Delta E \sim$ 5-40. The angular resolution is only limited by the size of the CCD pixels (0.492\arcsec). For on-axis imaging, $\sim$85\% of the encircled energy lies within $\sim$1.8\arcsec~of the central pixel\footnote{{\url{https://cxc.cfa.harvard.edu/proposer/POG/}}}. Among the three observations used in this work, ObsID. 5548 and 6312 are on-axis observations on PSR B1853+01 and its PWN (aimpoint on S3), whereas ObsID.1958 aimed at the center of the SNR W44, with a telescope roll angle of $\sim$ 285$^{\circ}$, with PSR B1853+01 falling on the S2 chip with an off-axis angle of about 7\arcmin.

We reduced the data using the \textit{Chandra} Interactive Analysis of Observations, CIAO (version 4.14.0)\footnote{\url{https://cxc.cfa.harvard.edu/ciao/}}, together with the calibration database CALDB version 4.9.6\footnote{\url{https://cxc.cfa.harvard.edu/caldb/}}. All three observations were reprocessed using \texttt{chandra$\_$repro} while for ObsID. 5548 and 6312, "\texttt{check$\_$vf$\_$pha = yes}" was set to reduce the ACIS particle background in very faint (VFAINT) mode observations.

Broad-band (0.5-7.0 keV) flux images accounting for the three $Chandra$ observations were produced using \texttt{fluximage} (with the parameter \texttt{psfecf}=0.9 to detect faint sources). These flux images were used later by \texttt{vtpdetect} to detect and exclude sources. We checked for flares by filtering the background light curve (0.5-7 keV, time bin width of 259.28 s) for deviations larger than 3$\sigma$ using CIAO's \texttt{deflare} tool. The output good time intervals were then used to filter the events file. No significant flares were detected in ObsID. 5548 and ObsID. 6312, accounting for unchanged exposure time of 52.31 ks and 39.58 ks, respectively, whereas for ObsID. 1958, there are only 44.79 ks left after background flares are removed. The total exposure for the three observations amounts to $\sim$ 136.86 ks. 

We also applied astrometric corrections to the data sets using the longest observation, ObsID. 5548, as the reference. We used \texttt{fluximage} (with parameter \texttt{psfecf}=0.393 to have bright point-like sources detected) to generate broad-band flux images that were later used by \texttt{wavdetect} to obtain source lists of each observation. After cross-matching the resulting lists obtained for ObsID. 1958 and ObsID. 5548 and then between ObsID. 5548 and ObsID. 6312 with \texttt{wcs$\_$match}, we corrected for the aspect solution of ObsID. 1958 and ObsID. 6312 with \texttt{wcs$\_$update}. After this astrometry correction, \texttt{merg$\_$obs} was then used to merge observations. For spectral analysis, \texttt{specextract} was used to extract spectra from individual observations while \texttt{grppha} was subsequently used for grouping the spectra. 

\subsection{\textit{NuSTAR}} 
The Nuclear Spectroscopic Telescope Array (\textit{NuSTAR}) observed the southern part of W44, containing PSR B1853+01 and its PWN, on Sepetember 2nd, 2018 (ObsID. 40401005002) for a total of $\sim$ 105 ks.

\textit{NuSTAR} is a focusing high-energy (3--79 keV) X-ray observatory \citep{Harrison2013}, featuring two co-aligned telescopes designed with conical Wolter-I type optics modules (OMA and OMB) that are paired with two solid state detector focal plane modules (FPMA and FPMB). With a focal length of $10.14$~m, \textit{NuSTAR}'s focal plane covers a FOV of $\sim$12\arcmin. \textit{NuSTAR} can reach sub-arcminute spatial resolution, with a PSF with an FWHM of 18\arcsec~and a half-power diameter of 58\arcsec, which is relatively constant across the FOV. It also has good spectral resolution, with an energy resolution (FWHM) of $\sim 400$~eV at 10~keV. 

\textit{NuSTAR} data used in this work were reduced using the NuSTARDAS package (version 2.1.1), along with \textit{NuSTAR} FPM caldb (version 20220105). \texttt{nupipeline} was used to reprocess the data to produce cleaned and calibrated event list files. For the imaging analysis, we first selected photon events in different energy bands using CIAO tool \texttt{dmcopy}. These energy bands were selected to ensure sufficient photon statistics in each image, as well as to minimize the effects of averaging the energy-dependent vignetting effect over the chosen energy range. We also used \texttt{nuexpomap} to produce exposure maps for each energy band, both for FPMA and FPMB. Then summing up \textit{NuSTAR} raw images in each energy band with these exposure maps, we created the final mosaic images using XIMAGE (HEAsoft 6.29). For spectral analysis, we used \texttt{nuproducts} to extract spectra for different regions in the FOV, with the parameter "\texttt{rungrppha=yes}" to generate grouped spectra. 
\begin{figure*}[!ht]
    \centering
     \includegraphics[width=\linewidth]{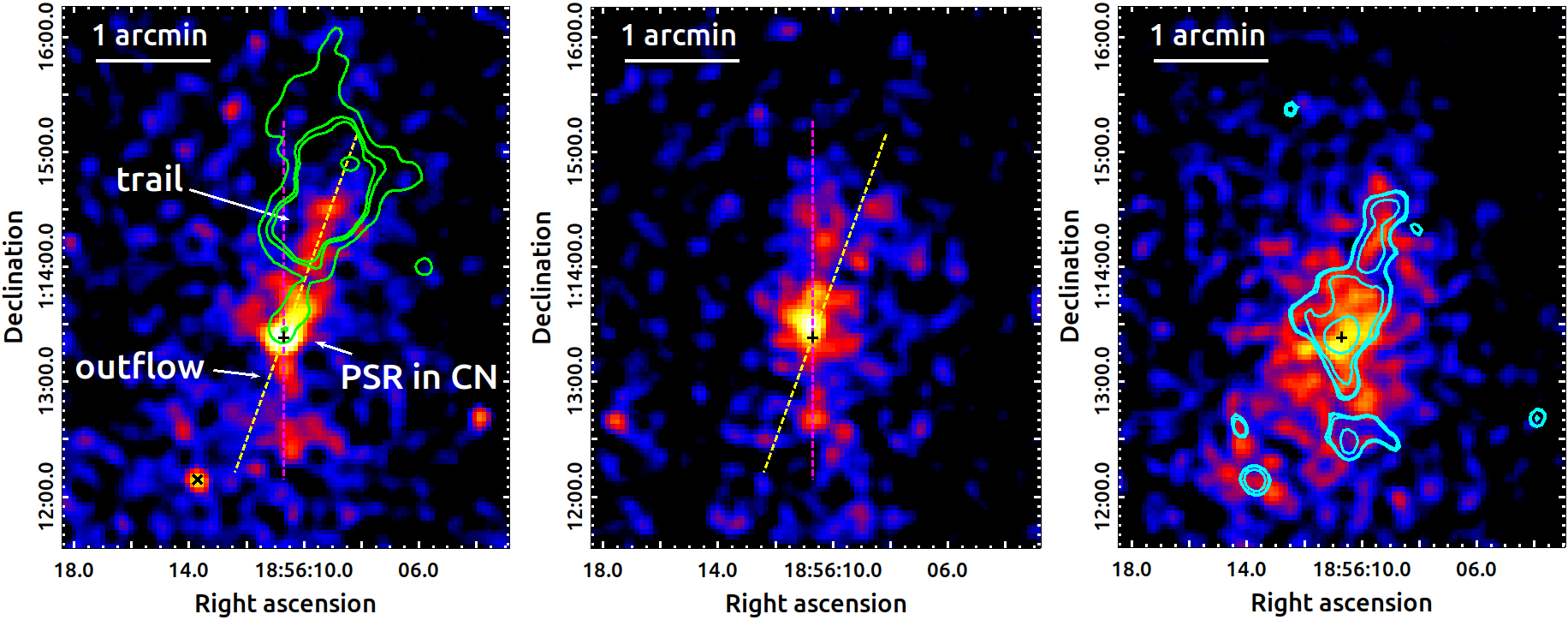}
    \caption[Caption for LOF]{Images of PSR B1853+01 (position marked by a black cross) and its surrounding emission. \textbf{Left}: Exposure-corrected merged \textit{Chandra} ACIS image in the 4-8 keV energy band, created by combining all three archived observations. The green contours show the radio emission from the PWN (radio FITS file obtained from SNRCat\protect\footnotemark). The image is binned by a factor 5 (pixel size=2.46"). The dashed yellow line shows the approximate symmetry axis of the PWN tail, which can be divide into two subregions: a compact nebula (CN) surrounding the pulsar and a trail emission further out. The dashed magenta line indicates the N-S direction which roughly coincides with the direction of the outflow just leaving the PWN bow shock. The position of a nearby source 2CXO J185613.6+011207 is marked by a black X. \textbf{Middle}: Merged (MOS1+MOS2) mosaic on-axis (Obs. 0551060101) \textit{XMM-Newton} counts image in the 4-8 keV energy band (pixel size=2.5"). \textbf{Right}: Merged (FPMA+FPMB) mosaic \textit{NuSTAR} image in the 4-10 keV band (pixel size=2.46"), together with the contour map (in cyan) extracted from the \textit{Chandra} image shown in the left panel. All images are smoothed with a Gaussian kernel of $r$=3~pixels and $\sigma$=1.5~pixels}.
    \label{fig: core nebula}
\end{figure*}

\begin{figure*}[!h]
    \centering
    \includegraphics[width=0.99\textwidth]{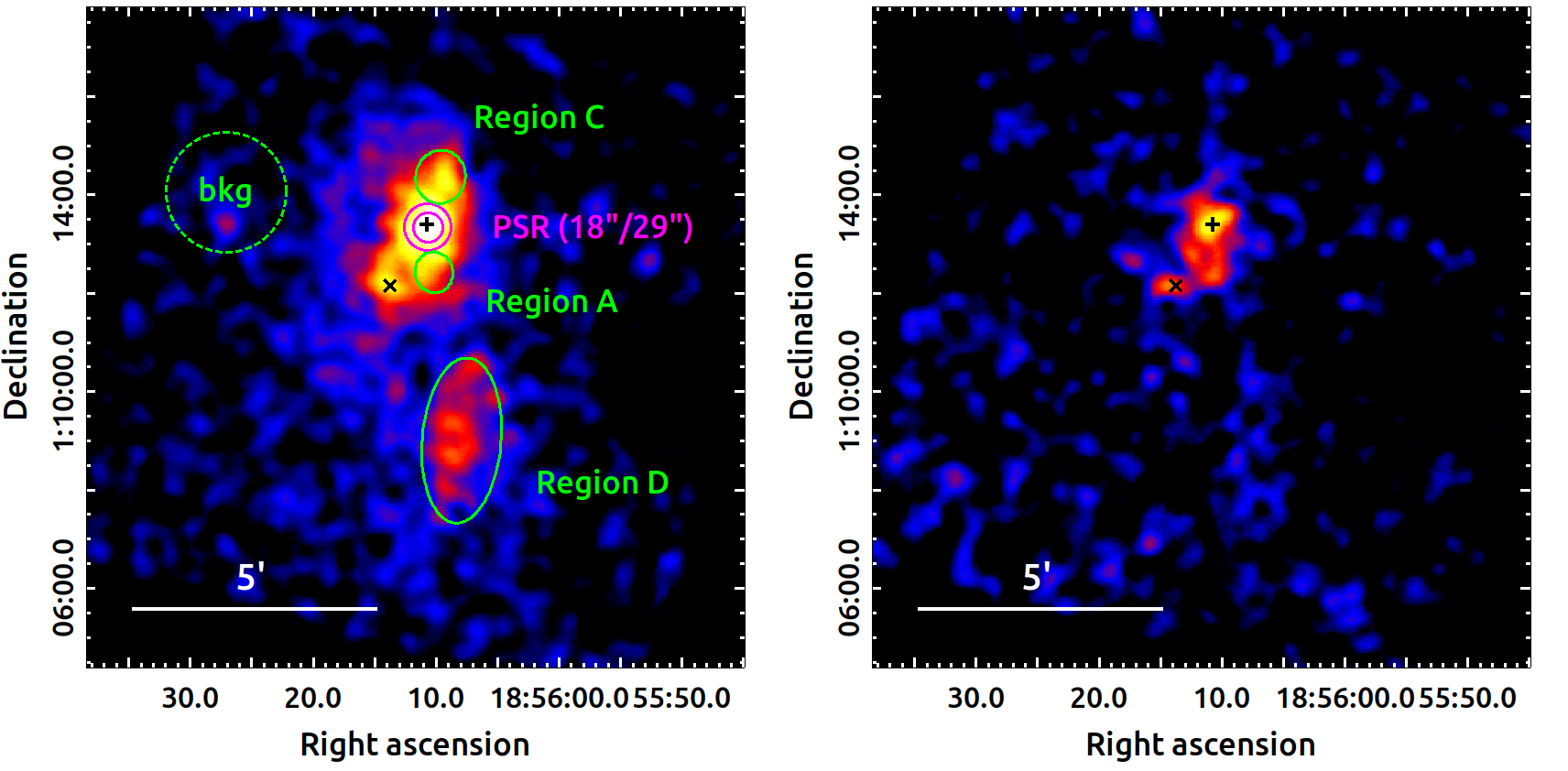}
    \caption{Merged (FPMA+FPMB) \textit{NuSTAR} images obtained in increasing energy bands. Images are smoothed with a Gaussian kernel of $r$=8~pixels and $\sigma$=4~pixels. Position of PSR B1853+01 is marked by a black cross while the position of a nearby source 2CXO J185613.6+011207 is marked by a black X. \textbf{Left}: The 4.0-10.0 keV image shows an extended emission coincides with the position of PSR B1853+01, exhibiting a spindle-like shape tapering at both ends and a faint extended emission in the south. Regions used for spatially resolved spectral analysis are shown here. Pulsar spectra were tried to extract with magenta circles of two different radii 18"/29". Region A is similar to the outflow region, Region C is similar to the trail region and D is similar to part of the southern halo where one of the large antenna feature lies. Background region is shown in dashed green circle. \textbf{Right}: The 10.0-20.0 keV image. The PSR+PWN system is detected up to energies of $\sim 20$ keV.}
    \label{fig:NuSTARimges}
\end{figure*}

\section{Analysis and Results}
The analysis of the data obtained with \textit{XMM-Newton}, \textit{Chandra} and \textit{NuSTAR} as described above provides complementary information on \PSRB\ and its nebula accounting for the unique strengths of each instrument. \textit{Chandra} offers an excellent angular resolution, which allowed us to separate the pulsar emission from that coming from the surrounding PWN and to resolve out structures at small scales. \textit{XMM-Newton}, featuring a unique sensitivity and a relatively large FOV, can be used to access the large-scale structures of low surface brightness around the PWN. Finally, the analysis of the \textit{NuSTAR} data is used to constrain the emission at hard X-rays, revealing for the first time the PWN in X-rays at energies above 10 keV. Below, we detail the obtained imaging and spectral analysis results.
\footnotetext{\url{http://snrcat.physics.umanitoba.ca/SNRrecord.php?id=G034.7m00.4}}

\subsection{Imaging}\label{imaginganalysis}

The PWN associated with PSR B1853+01 is hard to resolve at soft X-ray energies. However, it can be discerned from the surrounding SNR above 2.0 keV and becomes the dominant feature in the southern part of W44 above 4 keV (see Fig.~\ref{fig:W44}). We therefore limit our analysis of the PWN to relatively high energies in order to remove the contamination from the SNR thermal emission (a strategy employed by \citet{Harrus1996} and adopted also by \citet{Petre2002ApJ}).

Fig.~\ref{fig: core nebula} presents X-ray images of PWN B1853+01---the region immediately surrounding PSR B1853+01---obtained with \textit{Chandra} (4-8 keV), \textit{XMM-Newton} (4-8 keV) and \textit{NuSTAR} (4-10 keV). Here, we selected similar energy bands and also made efforts to closely match the pixel size and smoothing strategy across the three instruments, to ensure a consistent characterization of this region. In particular, the \textit{XMM-Newton} counts map was produced using only MOS data from the on-axis observation Obs. 0551060101, as this selection provided the best resolution \textit{XMM-Newton} could achieve. The \textit{Chandra} and \textit{XMM-Newton} images reveal complex diffuse X-ray structures surrounding the pulsar: an extended emission around the position of the pulsar stretching in the NE-SW direction, a trail of emission extending further to the northwest, and in addition, an extended emission detected to the south of the pulsar. On larger scales beyond the zoomed-in field shown in Fig.~\ref{fig: core nebula}, a faint, halo-like, diffuse emission encompassing the PWN is clearly resolved by \textit{XMM-Newton} as shown in Fig. \ref{fig:W44}.

Full-frame \textit{NuSTAR} images merging two FPMs were created in the energy bands: 4-10 keV, 10-20 keV and 20-79 keV. Due to the absence of significant source detection in the 20–79 keV range, we show only the 4–10 keV and 10–20 keV \textit{NuSTAR} images in Fig.~\ref{fig:NuSTARimges}. We note that this is the first detection of this PWN above 10 keV, and reaching $\sim$20 keV. Moreover, in the 4-10 keV \textit{NuSTAR} image, a faint extended feature is found at the same location as the large-scale structure revealed in the \textit{XMM-Newton} images south of the PWN.

\begin{figure}[!ht]
    \includegraphics[width=0.95\linewidth]{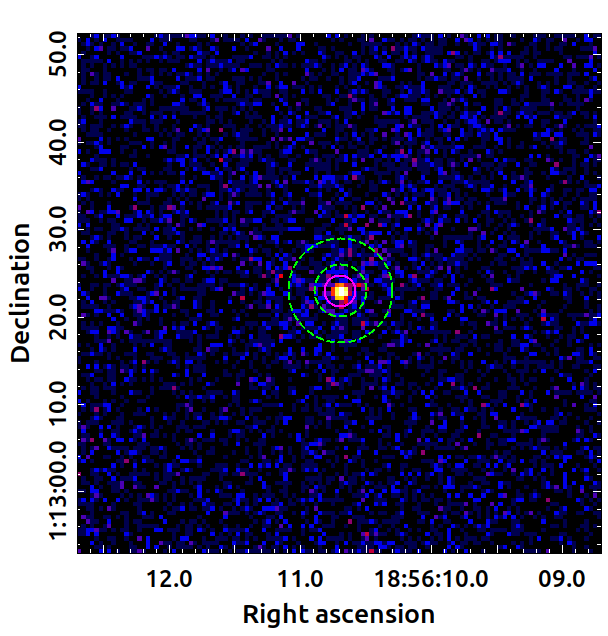}\caption{\textit{Chandra} 2-7 keV on-axis counts image, unbinned and unsmoothed. Circle and annulus are centered at the pulsar position detected by \textit{Chandra}. The magenta circle as a radius of 1.8\arcsec and is used for the extraction of the pulsar spectra; the green dashed annulus, with inner/outer radii of 3\arcsec and 6\arcsec, is used as the background region.}
    \label{fig:chandraonaxisunbinnedhard_countimage}
\end{figure}

\subsubsection{The Pulsar B1853+01}
Between the first off-axis (obs. 1958) and the later two on-axis (obs. 5548,6312) \textit{Chandra} observations, about 4 years and 7 months, we detect no significant shift of pulsar position. The two on-axis \textit{Chandra} observations (Obs. 5548 and 6312) were used to further study the compact source. An unbinned and unsmoothed count image in the \textit{Chandra} hard energy band (2.0-7.0 keV) made by combining the two observations is shown in Fig. \ref{fig:chandraonaxisunbinnedhard_countimage}. In this energy range, the emission from the pulsar dominates that from the surrounding PWN and also that from the SNR thermal emission. We performed source detection on the combined counts image using \texttt{wavdetect} with both PSF and exposure maps supplied as inputs and pinpointed the pulsar position at R.A.(J2000)= 18\hours56\minutes10\fs68, Dec.(J2000)=+01\degrees13\arcmin22\farcs87. In Table~\ref{tab:pulsarposition}, we summarize the positions of PSR B1853+01 obtained from previous timing observations in the radio band, from previous X-ray measurements as well as from our own work. The errors given in the \textit{Chandra} measurements include both statistical and systematic uncertainties. A comparison of \textit{Chandra} and radio positions implies a transverse velocity significantly higher than that estimated by \citet{Frail1996ApJ} based on the position of the pulsar within the SNR. However, this discrepancy may arise from underestimated uncertainties in radio position reported by \citet{Jacoby1999}, as PSR B1853+01 exhibits substantial intrinsic timing noise. It is noticed that in the \textit{XMM-Newton} MOS image of the region surrounding PSR B1853+01 created from Obs. 0551060101, the centroid of the brightest pixel is located $\sim$ 4\arcsec~north to the position determined as mentioned above from \textit{Chandra} on-axis observation data. We thus further examine the images produced from Obs. 721630101, 721630201 and 721630301, and found that the centroids of the brightest pixels in these images coincide with the \textit{Chandra} pulsar position. We conclude that a relatively small position offset is present in Obs. 0551060101. On the other hand, the centroid of the brightest pixels in the 4.0-20.0 keV \textit{NuSTAR} mosaic image is $\sim$6\arcsec~away from the derived \textit{Chandra} pulsar position, yet it is within \textit{NuSTAR}'s absolute astrometric uncertainty of 8" for the brightest targets (90\% confidence; \citet{Harrison2013}).

\begin{table}[h!]
    \centering
    \caption {Position of PSR B1853+01.}
    \renewcommand{\arraystretch}{1.5}
    \begin{tabular}{c c c}
    \toprule
    \toprule
        &R.A. (J2000.0) & Decl. (J2000.0)\\
    \midrule    
        Radio$^a$ & 18 56 10.76 $\pm$ 0.02 & 01 13 28.0 $\pm$ 0.6\\
        ASCA (GIS)$^b$ & 18 56 10.0 $\pm$ 1.5 & 01 13 17 $\pm$40 \\
        ASCA (SIS)$^b$ & 18 56 12.6 $\pm$1 & 01 12 34 $\pm$ 30\\
        \textit{Chandra}$^c$ & 18 56 10.69 $\pm$ 0.03 & 01 13 22.87 $\pm$ 0.49\\
        \bottomrule
    \end{tabular}
    \tablefoot{
    \tablefoottext{a}{from \citet{Jacoby1999}.}
    \tablefoottext{b}{from \citet{Harrus1996}; celestial coordinates are assumed to correspond to the central peak of the X-ray PWN.}
    \tablefoottext{c}{from this work; position obtained by applying CIAO tool \texttt{wavdetect} to \textit{Chandra} on-axis observation data corresponding to Obs. 5548 and Obs. 6613 taken on June 25th 2005 and June 23rd 2005, respectively, accounting for both statistical and systematic 68\% C.L. uncertainties.}
    }
    \label{tab:pulsarposition}
\end{table}

\subsubsection{The PWN Tail}\label{tail}
The merged \textit{Chandra} image (Fig.~\ref{fig: core nebula}, left panel) reveals a cometary-shaped "tail" emission, which is elongated in the southeast-northwest direction, approximately along a line with a position angle of $\sim$20\degree~(north through west). This tail feature, extending up to $\sim$1.5\arcmin~(or 1.4 pc at $d =$ 3.2 kpc) on the northern side of the pulsar, is located inside of the thermal emission boundary of the SNR and is about half the extent observed at radio wavelengths. A gap half way along the tail is clearly visible in the \textit{Chandra} image (also in \textit{XMM-Newton} and \textit{NuSTAR} images), dividing the tail morphologically into two subregions: a ``near'' half identified as the compact nebula (CN) in the pulsar's very vicinity, and a ``far'' half refereed to as the ``trail'' region. The CN is the second brightest feature in the field, about ten times fainter than the pulsar, whereas the extended trail emission is relatively dimmer with an average surface brightness approximately half that of the CN. 
A hint of a small extended feature, at an angle to the pulsar's proper motion direction (protruding to the northeast just behind the pulsar), of similar brightness to the ``trail'' region, is also apparent in the \textit{Chandra} image and renders the tail emission less symmetric.

\subsubsection{The Outflow}  
 
We identify an extended region of emission, the ``outflow'', preceding the pulsar with a surface brightness similar to that of the ``trail'' region. This emission region, which lies mostly outside of the thermal emission boundary of the SNR, extends as far as $\sim$1\arcmin~(or $\sim$1.0 pc at $d =$ 3.2 kpc) to the south of the pulsar and covers approximately 1000 $\textrm{arcsec}^2$ in the sky. A total of 3202 source counts were retrieved from this outflow region, whereas about 2522 background counts are estimated, yielding 680 excess counts, implying a $\sim9\sigma$ statistical significance assuming a Gaussian approximation. Best resolved by \textit{Chandra}, this outflow feature exhibits a morphology resembling a pair of "antennae". The 4.0-8.0 keV \textit{Chandra} image in Figure \ref{fig: core nebula} shows an extended emission that protrudes from the southern vicinity of the pulsar, subsequently breaking into two filamentary branches while a gap in the emission is discernible halfway before the branching. One of the southern branches continues in the south direction, whereas the other extends towards the southwest direction. \textit{XMM-Newton}, on the other hand, can barely resolve this extended feature from the CN+PSR emission, yet it detects the gap between the protruding stem and the southern branching part. These findings are in good agreement with the \textit{Chandra} image presented in \citet{Petre2002ApJ}, where hints of this extended region of emission appeared as part of a fainter subregion used for spectral analysis there.

\subsubsection{An X-ray Halo}\label{haloimageanalysis}
A relatively fainter yet more extended emission, encompassing the central PWN is discernible in \textit{XMM-Newton} image in Fig. \ref{fig:W44} and Fig. \ref{fig:xmmnewtonextraction}. This extended emission, which we tentatively identify as an X-ray ``halo'', displays an average surface brightness dimmer than that of the PWN by a factor $\sim$3. The halo region can be further divided into the northern and southern parts based on the boundary of the thermal SNR X-ray emission. The northern part of the X-ray halo follows the overall shape of the PWN tail, extending only $\sim$ 0.5 arcmin outward from the PWN. In contrast, the southern part has a diffuse large-scale antennae-like morphology, reaching distances of about 6\arcmin/3\arcmin~(or 6.0/3.0 pc at d= 3.2 \textrm{kpc}). This prominent large-scale southern halo feature appears to extend from the central PWN, in the very same direction as the smaller and brighter inner feature\textemdash the outflow\textemdash observed at closer distances south to the pulsar.

Our results from \textit{NuSTAR} confirm the existence of the large-scale structure (left panel in Fig.~\ref{fig:NuSTARimges}) with a morphology closely resemble that obtained with \textit{XMM-Newton}, despite possible distortion caused by chip gaps in the \textit{NuSTAR} data. We also searched for this large-scale structure in \textit{Chandra} images. This was prevented, however, by the orientation of the ACIS array, as the two on-axis observations failed to cover this area. Only in Obs.ID 1958 is this region covered, but most of the feature extension falls on the edge of the S2 chip and the back-illuminated S1 chip. 

\begin{table*}[!ht]
    \newcolumntype{C}[1]{>{\centering\arraybackslash}p{#1}}
    \centering
    \caption{Best-fitting values for PSR B1853+01 }
    \begin{threeparttable}
    \renewcommand{\arraystretch}{1.5}
    \begin{tabular}{p{3.25cm}C{3.25cm}C{3.2cm}C{3.2cm}C{3.2cm}}
    \toprule
    \toprule
     & $\mathrm{N}_{\mathrm{H}}$ ($\times 10^{22}~\textrm{cm}^{-2}$) &  $\Gamma$ &  $\textrm{Norm}^a$ & $\chi^2_{\textrm{red}}~\textrm{(d.o.f)}$\\
    \midrule
    \multicolumn{5}{l}{{\tt{tbabs} $\times$ PL:}}\\
    \textit{Chandra}& $0.65^{+0.46}_{-0.42}$ & $1.87^{+0.48}_{-0.43}$ & $1.22^{+0.86}_{-0.49}\times10^{-5}$ & 1.01~(16)  \\ 
    \textit{XMM-Newton} & $0.75^{+0.17}_{-0.14}$ & $1.70^{+0.17}_{-0.16}$ & $1.45^{+0.35}_{-0.28}\times10^{-5}$& 0.97~(248) \\ 
    \textit{NuSTAR}$^b$ &  0.65 (fixed) & $1.91^{+0.42}_{-0.43}$ & $8.24^{+10.37}_{-5.39}\times10^{-5}$&  1.20~(19) \\ 
    \bottomrule
    \multicolumn{5}{l}{\footnotesize \textbf{Note}: $^a$ photons/keV/cm$^2$/s @ 1 keV. $^b$value for FPMA; for FPMB Norm is $9.83^{+8.76}_{-4.52}\times10^{-5}$}
    \end{tabular}
    \end{threeparttable}
    \label{tab:psrspecfitting}
\end{table*}

\subsection{Spectral Results}

In our spectral analysis, spectra were extracted from the following regions: (1) the pulsar region, (2) the PWN tail region (CN+trail; excluding the pulsar itself), (3) the ``outflow'' region ahead of the pulsar, and (4) the faint ``halo'' region encompassing the three aforementioned regions. We also defined several background regions to be used in the spectral fits. The XSPEC \citep{Arnaud1996ASPC} software (version 12.9.0) was used for the spectral analysis. The Tuebingen-Boulder ISM absorption model \texttt{tbabs} with abundances taken from \citet{Wilms2000ApJ} were used to calculate the X-ray absorption in the ISM. Uncertainties are given at 90$\%$ confidence level unless specified otherwise.

\subsubsection{The Pulsar B1853+01}

The thermal contamination from the SNR in the soft X-ray band (below $\sim$4 keV) significantly impacts the spectral analysis of the whole PWN. The pulsar itself might be in this regard the less affected region given its point-like nature. The spectral analysis on PSR B1853+01 has been therefore used to determine the column density, $\mathrm{N}_{\mathrm{H}}$, towards the system. 

For the analysis of \textit{Chandra} data, a circular region with 1.8" radius centered at $\textrm{R.A.(2000)}=18\hours56\minutes10.69\seconds$, $\textrm{Dec.(2000)}=01\degrees 13\arcmin 22.9\arcsec$ was defined as the extraction region for PSR B1853+01. To remove the contamination from the PWN, the background region was chosen to be a $3\arcsec-6\arcsec$ annulus centered on PSR B1853+01 (see Fig. \ref{fig:chandraonaxisunbinnedhard_countimage}). The spectra were extracted from the three \textit{Chandra} observations and were combined and grouped to have a minimum of 20 cts/bin. We then fitted the pulsar spectra from 0.5 to 8 keV using different models. By applying an absorbed power-law emission model, ``\texttt{tbabs} $\times$ PL'', the best fit yields a column density of $\mathrm{N}_{\mathrm{H}}$ = $ 0.65^{+0.46}_{-0.42} \times 10^{22}$ cm$^{-2}$, a photon index of $\Gamma$=$ 1.87^{+0.48}_{-0.43}$, a normalization of $1.22^{+0.86}_{-0.49}\times10^{-5}$ $\textrm{photons/keV/cm}^{2}\textrm{/s}$ and an observed flux in the 0.5-8.0 keV band of $4.12 \times 10^{-14}$~ergs~cm$^{-2}$~s$^{-1}$ ($\chi^2_{\textrm{red}}$ = 1.01 with 16 degrees of freedom, null hypothesis probability of 0.44). We note also that middle-aged pulsar often have thermal component the pulsar spectrum. However, fitting the pulsar spectra with an absorbed blackbody model alone (``\texttt{tbabs} $\times$ BB'') gives an unrealistically low column density with a large associated error. Parameters could not be well constrained neither when fitting the pulsar spectra with an absorbed blackbody plus a power law emission model, ``\texttt{tbabs} $\times$ (BB + PL)''. We also tried to extract the pulsar spectra with other different background regions drawn from nearby fields. Fit results were insensitive to the specific choice of background region. Finally, as a further check, we considered also deriving the $\mathrm{N}_{\mathrm{H}}$ value from the CN region, the spectrum of which is expected to be a pure power-law only modified by the ISM absorption. The results of the spectral fit are however highly sensitive to the choice of the background regions. Using the background region ``bkg1'' defined as in Figure 5, the best fit $\mathrm{N}_{\mathrm{H}}$ value from the CN region is $ 2.95^{+1.17}_{-0.88} \times 10^{22}$ cm$^{-2}$ ($\chi^2_{\textrm{red}}$ = 1.11 with 145 degrees of freedom, null hypothesis probability of 0.18).

Our best-fit pulsar photon index $\Gamma$=$ 1.87^{+0.48}_{-0.43}$ differs noticeably from the significantly harder spectrum, $\Gamma= 1.28\pm 0.48$, reported in \cite{Petre2002ApJ}. We noticed that these authors consider a much higher column density $\mathrm{N}_{\mathrm{H}} \approx 5\times10^{22}~\textrm{cm}^{-2}$, possibly referencing an incorrect value from \citet{Rho1994ApJ}, as the two sub-figures describing the temperature and column density distributions of W44 region by region in Fig. 10 of \citet{Rho1994ApJ} are wrongly labeled (according to the values in their Table 4). The value of $\mathrm{N}_{\mathrm{H}}$ reported here is, however, consistent with what obtained by \cite{Harrus2006ESASP} where the pulsar spectrum was extracted from a radius of 20" centered at $\textrm{R.A.(2000)} = 18\hours56\minutes11\seconds$, $\textrm{Dec.(2000)} = 01\degrees 13\arcmin 24\arcsec$, using \textit{XMM-Newton} Obs. 0083270301 and  0083270401. 
In \cite{Harrus2006ESASP}, the spectrum was best described with an absorbed power law model with $\Gamma =2.2^{+0.49}_{-0.42}$ using $\mathrm{N}_{\mathrm{H}}$ =$0.6\pm0.2\times10^{22}~\textrm{cm}^{-2}$, a column density value slightly lower than what they obtained for the entire remnant. \citet{Harrus1996}, on the other hand, presented ASCA analysis of the X-ray synchrotron nebula associated with PSR B1853+01 where data were extracted from a circular region of radius 5\arcmin(3\arcmin) centered on the radio pulsar position, obtaining a photon index $\Gamma = 2.3^{+1.1}_{-0.9}$ and a column density $\mathrm{N}_{\mathrm{H}} = (1.82\pm 0.05)\times 10^{22}~\textrm{cm}^{-2}$. The moderate discrepancy with our results might not be surprising as the extraction regions applied are much larger, in contrast to the \textit{Chandra} capabilities in spatially resolving the different structures.

For \textit{XMM-Newton} data, given the different PSFs of the detectors, we defined pulsar extraction regions of different sizes. For Obs.551060101, 
we used circular source regions of 7\arcsec/9\arcsec~(MOS/\textit{pn}) and concentric annuli background regions of 8\arcsec–11\arcsec (MOS) and 10\arcsec–13\arcsec (\textit{pn}).For other off-axis observations, source radii were 9\arcsec/11\arcsec~(MOS/\textit{pn}), with background annuli of 10\arcsec–13\arcsec~(MOS) and 12\arcsec–15\arcsec ~(\textit{pn}). Extracted spectra were combined and grouped to $\geq$30 cts/bin. Fitting the spectrum in 0.5–10 keV gives an absorbed power law with $\mathrm{N}_{\mathrm{H}}$ = $0.75^{+0.17}_{-0.14}\times 10^{22}\textrm{cm}^{-2}$ and $\Gamma = 1.70^{+0.17}_{-0.16}$ (see Table~\ref{tab:psrspecfitting}), consistent, within errors, with \textit{Chandra} results. In the remainder of the spectral analysis, $\mathrm{N}_{\mathrm{H}}$ = 0.65$\times 10^{22}$ cm$^{-2}$ will be used. For \textit{NuSTAR} data, pulsar spectra extracted from an 18\arcsec/29\arcsec~circular region fixed at the centroid of the brightest pixels were grouped to have at $\geq$ 200 cts/bin. When applying an absorbed power-law model with $\mathrm{N}_{\mathrm{H}}$ fixed, we get a best fit in 3-10 keV with $\Gamma=1.91^{+0.42}_{-0.43}$, $\chi^2_{\textrm{red}}\textrm{ (d.o.f)}$ of 1.20 (15) (see Table~\ref{tab:psrspecfitting})/ $\Gamma=2.20^{+0.27}_{-0.27}$, $\chi^2_{\textrm{red}}\textrm{ (d.o.f)}$ of 0.75 (36).

\subsubsection{Spatially Resolved Spectral Analysis}\label{spectralresults}

Spatially-resolved spectra of the extended emissions surrounding the pulsar were extracted from different regions in \textit{Chandra}, \textit{XMM-Newton}, and \textit{NuSTAR} data sets.
\begin{figure}[!ht]
    \centering
    \includegraphics[width=1.0\linewidth]{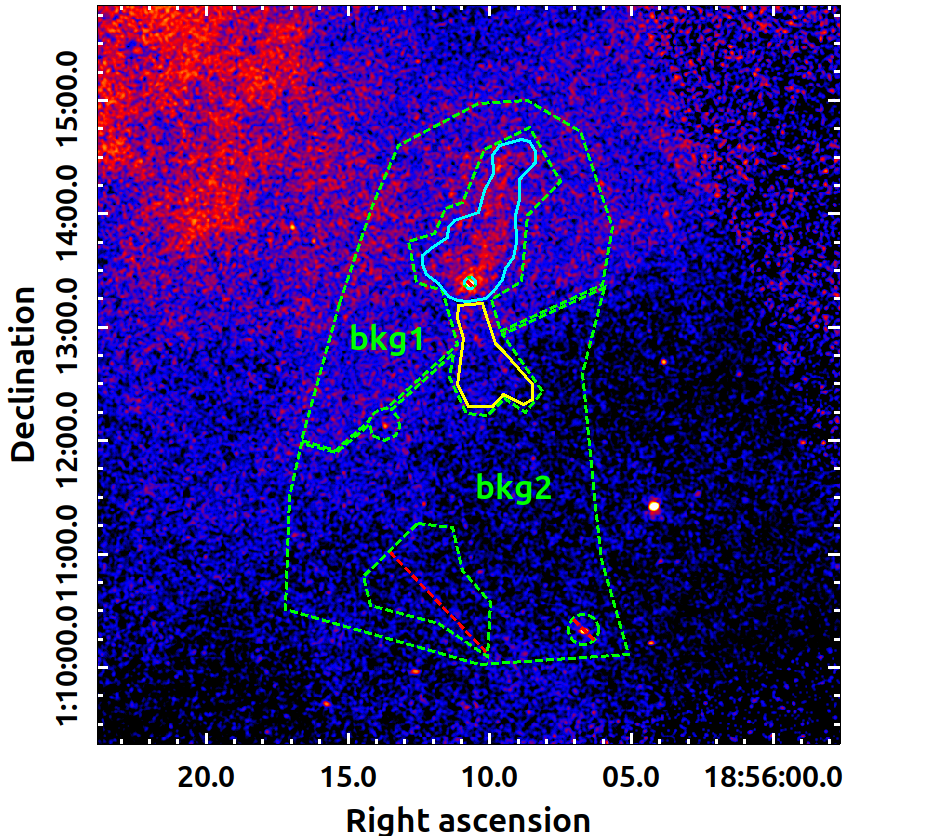}
    \caption{Merged \textit{Chandra} broad band (0.5-7.0 keV) flux image with extraction regions used for spectral extraction overlaid. The image is unbinned yet smoothed with a Gaussian function with radius r=3 pixel and sigma $\sigma$=1.5 image pixels. Weighted spectra have been extracted from tail (solid cyan) region with pulsar (r=3") excluded, using bkg1 region (dashed green) as background region, and from outflow region (solid yellow) using bkg2 region (dashed green) as background region.}
    \label{fig:PSRB1853chandrapwnspec}
\end{figure}

For the three archived \textit{Chandra} observations, weighted spectra were extracted from the PWN tail region (CN+trail) and the outflow region, as shown in Fig. \ref{fig:PSRB1853chandrapwnspec}. The imaging analysis results in Sec. \ref{imaginganalysis} reveals that the PWN tail is located inside the boundary of the SNR thermal X-ray emission while most of the outflow region lies outside and the whole PWN is surrounded by a relatively fainter halo-like structure (see Sec. \ref{haloimageanalysis}). 
For the PWN tail region, we chose a background region (denoted as "bkg1" in Fig. \ref{fig:PSRB1853chandrapwnspec}) similar to the part of the halo structure inside the boundary of the SNR thermal X-ray emission. For the outflow feature a background region was taken from the halo structure outside of the boundary of the SNR thermal X-ray emission but still enclosing the outflow region (denoted as "bkg2" in Fig. \ref{fig:PSRB1853chandrapwnspec}). The weighted spectra extracted from the two regions were combined and grouped respectively to have at least 20 cts/bin and fitted with an absorbed power-law model, ``\texttt{tbabs} $\times$ PL''. In Table \ref{tab:spectral} we present the best fit results from fitting the spectra extracted from the two regions. The tail spectra were fitted from 3.0 to 8.0 keV (to minimize the effect of the residual thermal emission) with the column density fixed at $0.65\times 10^{22}$~cm$^{-2}$, yielding a photon index of $\Gamma = 2.01^{+0.39}_{-0.38} $, $\chi^2_{\rm {red}}$= 0.98 (d.o.f.=53) for this tail region. We tried to extract and fit the spectra from the two subregions (CN and ``trail'') of the tail structure, but the low statistics restricted us from obtaining meaningful results after cutting out the thermal radiation below 3 keV. The spectra extracted from the outflow region were fitted on the other hand from 0.5 to 8.0 keV and the best fit results give a rather hard photon index, $\Gamma=1.24^{+0.23}_{-0.24}$. 

\begin{figure*}[!ht]
 \centering
    \includegraphics[width=1.0\linewidth]{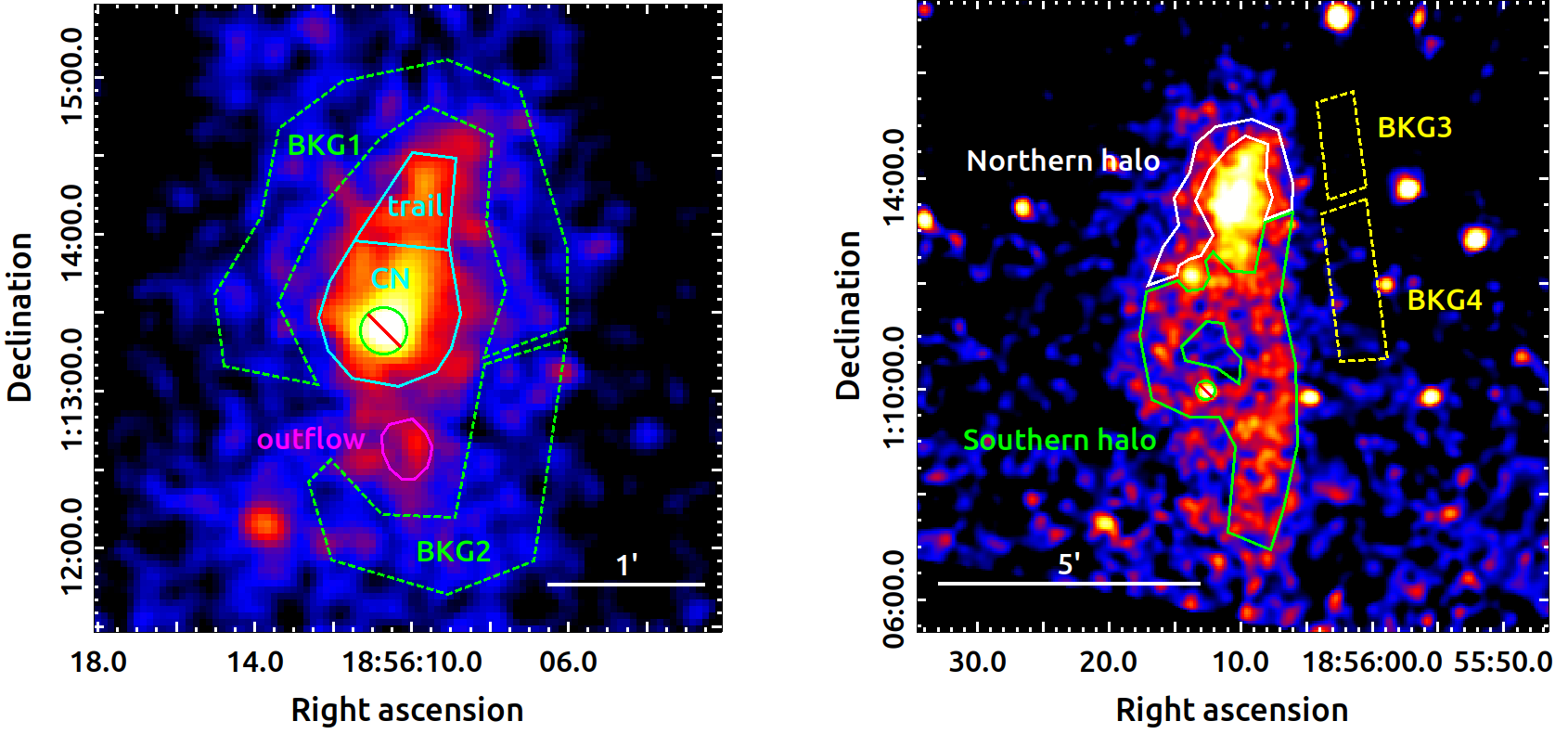}
    \caption{\textbf{Left}: 4\arcmin$\times$4\arcmin~\textit{XMM-Newton} \textit{pn} image in 4.0-8.0 keV band (merged from Obs. 0721630101, 0721630201, and 0721630301). The extraction regions for the tail region (solid cyan; crossed green circle region representing PSR B1853+01 was excluded from the CN/tail spectral analysis) and the outflow region (solid magenta) are shown. The tail region can be further split into two subregions: CN and trail (separated by the line). A background region inside the thermal X-ray emission area BKG1 (dashed green) was chosen for the tail regions while one outside BKG2 (dashed green) for the outflow. \textbf{Right}: 12\arcmin$\times$12\arcmin~\textit{XMM-Newton} \textit{pn} image in 4.0-8.0 keV band (merged from Obs. 0721630101, 0721630201, and 0721630301).The extraction regions for the halo region are shown. The halo region is divided into two subregions: the northern halo (solid white) and the southern halo (solid green). Background regions (dashed yellow lines) for these two regions of interest are BKG3 and BKG4 respectively. The point source (shown by the crossed green circle) that appears in the southern halo region was excluded from the halo spectral analysis.}
    \label{fig:xmmnewtonextraction}        
\end{figure*}

 For \textit{XMM-Newton} observations, we use only \textit{pn} data from Obs. 721630101, 0721630201 and 0721630301. Despite the relative worse spatial resolution of the \textit{pn} camera, these exposures were chosen as the entire PWN $+$ halo system was captured on one single \textit{pn} chip. In contrast, other EPIC exposures are compromised and not used for either significant portion of the emission is not available (due to the loss of MOS1 CCD3, which affected Obs ID.721630101, 0721630201 and 0721630301), or the PWN falls in close proximity to chip gaps or bad pixels. We extracted spectra from the PWN tail region, the outflow region and also the halo region as shown in Fig.~\ref{fig:xmmnewtonextraction}. For the PWN tail and outflow region, background regions are selected in nearby halo regions as shown in the left panel of Fig.~\ref{fig:xmmnewtonextraction}. Extracted spectra were grouped and combined to have at least 30 cts/bin and were fitted using an absorbed power-law model, ``\texttt{tbabs} $\times$ PL'', with $\mathrm{N}_{\mathrm{H}}$ fixed to 0.65$\times10^{22}$~cm$^{-2}$. Specially for the PWN tail, the high sensitivity of \textit{XMM-Newton} allowed us to split it into two subregions: the CN region and the trail region. For both subregions, we fit the data in the energy range 3.0 to 10 keV as the low energy part is heavily affected by background subtraction. For the outflow region, since it is outside of the SNR thermal X-ray emission, its spectrum is less affected in the low energy part and hence we fit it from 0.5 to 10.0 keV. A summary of the spectral fitting results is given in Table \ref{tab:spectral}. 
 We also tried to extract spectra from the halo region and its two subregions as shown in right panel of Fig.~\ref{fig:xmmnewtonextraction}. Spectra from the two subregions were combined and grouped to have at least 200 cts/bin due to the low surface brightness of this halo feature. We limit the fit from 3.0 keV to 7.5 keV to the halo spectra as they are rather noisy below 3 keV and copper line emission due to the detector itself is present in the high energy end outside of the fit range.

For \textit{NuSTAR} observations, spectra have been extracted from extended regions A, C and D as shown in Fig.~\ref{fig:NuSTARimges}. Region A is similar to the outflow region, region C is similar to the trail region and region D is similar to the part of the southern halo where one of the large-scale antenna feature lies. We fit the \textit{NuSTAR} spectra in the 3 to 20 keV band with an absorbed power-law model, ``\texttt{tbabs} $\times$ PL'', fixing the column density value at $N_{\textrm H}$ = 0.65$\times 10^{22}$cm$^{-2}$. The photon index parameter is linked for the two detector modules, whereas normalization parameters were allowed to vary to take into account the small relative calibration uncertainties between the two modules. Our best fit results are presented in Table.~\ref{tab:spectral}.

\section{Discussion}

\subsection{BSPWN B1853+01 inside of SNR W44}

In general, the evolution of the PWN is coupled to that of its associated SNR until either the pulsar overtakes the blast wave (at crossing time $t_{cr}$) or the SNR dissipates. Born with a given kick velocity, a pulsar powers a PWN that is initially expanding supersonically into the freely expanding SNR ejecta. When the PWN shock interacts with the SN reverse shock, the PWN bubble oscillates back and forth due to the reverberations from the passage of the reverse shock. Later, the reverberations vanishes and the PWN expands subsonically. Eventually, when the pulsar is positioned at $\sim 2/3$ of the radius of the supernova remnant blast wave, a bow shock is formed around the head of the PWN due to the mildly supersonic motion of the pulsar \citep{vanderSwaluw2004A&A}, which might be the case here for BSPWN B1853+01 inside the SNR W44. Following the formation of the bow shock, the Mach number will slowly increase following the decrease of the sound speed as the PWN approaches the SNR Shell (see e.g. the location of the BSPWN in SNR CTB80, close to the SNR shell;  \citealp{Safiharb1995, Migliazzo2002ApJ, Li2005ApJ}). Once the pulsar breaks through the shell, its dynamics are no longer affected by the parent SNR whereas it will continue being accompanied with a BSPWN for its supersonic motion in the ISM. BSPWNe are thus usually formed during the last stages of PWN/SNR evolution. 

PSR B1853+01 is estimated to be moving to the south with a transverse velocity $\lesssim$370 km s$^{-1}$ ($\pm$ 25 km s$^{-1}$) for the pulsar is displaced about 8\arcmin.6 $\pm$ 0\arcmin.5 away from the SNR center and the age for the SNR/PSR system is $\tau_c~\sim$20 kyrs \citep{Frail1996ApJ}. The tail that we observed in X-rays extending about $\sim$1.5\arcmin~ to the north confirms that the pulsar is indeed moving supersonically to the south. The pulsar wind is confined by the ram pressure, $\dot{E} / 4\pi R_{TS}^2 c= \rho_0 v_{PSR}^2$ ($R_{TS}$ is the stand-off distance, c is the speed of light, $\rho_0$ is the density of the ambient medium and $v_{PSR}$ is pulsar's velocity relative to the medium). Compared to other BSPWNe moving in the ISM, PWN B1853+01 has a rather compact size, with the tail extending only for about 1.4 pc. This is probably due to the existence of the high thermal pressure of the hot X-ray emitting gas in W44, $P_{\textrm{plasma}}\sim 1.1 \times$ 10$^{-9}$ ergs cm$^{-3}$, several orders of magnitude higher than the pressure of the ISM \citep{Harrus1997ApJ}. The X-ray tail emission is likely due to synchrotron radiation from the shocked pulsar wind. The size of this X-ray tail is about half the extent observed in radio is probably due to a shorter synchrotron loss timescale for X-ray emitting electrons. The gap detected in the tail region in our X-ray imaging analysis, corresponding approximately to the ``kink'' feature observed in radio observations \citep{Frail1996ApJ}, could possibly be produced due to the passage of the reverse shock. The photon index of the tail region ($\Gamma_{Chandra}=2.01^{+0.39}_{-0.38}$) is consistent with those of a typical PWN, which have nominal ranges of $\Gamma \sim 1.1 - 2.4 $ \citep{Gaensler2006ApJ}.

The \textit{Chandra} imaging results in Sec.~\ref{tail} shows that the angular separation from the pulsar position to the edge of the extended CN in the direction of the pulsar motion is about 8\arcsec. However the true location of the termination shock is expected to be much closer to the pulsar than the edge of the CN. In the case of the Crab nebula, for example, it is located at $\sim 10 \%$ of the outer radius of the luminous nebula. Assuming that the pulsar is located 3.2 kpc away and is moving in the plane of the sky and taking 10$\%$ of this as the standoff separation angular distance it yields an estimated standoff radius, $R_{TS}$, of approximately 0.01 pc. For a particle density $n_e = 0.25 ~\textrm{cm}^{-3}$ near PSR B1853+01 inside W44 \citep{Jones1993MNRAS}, assuming cosmic abundances, this corresponds to a mass density of $\sim 1.2 \times 10^{-24}~\textrm{g/cm}^3$.
The velocity of the pulsar is thus $ v_{PSR}= (\dot{E} / 4\pi \rho_0 R_{TS}^2)^{1/2}\sim 320$ km/s. Our estimate is consistent with that inferred from its relative position to the center of the SNR and its age. However, in both estimations, only the transverse component is accounted for. A precise measurement of the full 3D velocity would require observations with current radio interferometer facilities like VLA and MeerKAT.

\begin{table*}[!t]
\newcolumntype{C}[1]{>{\centering\arraybackslash}p{#1}}
    \centering
 \renewcommand{\arraystretch}{1.7}
    \caption{Best fit results from spatially resolved spectra}
    \begin{tabular}{p{2.4cm}p{2.4cm}C{2.4cm}C{3cm}C{3cm}C{3cm}}
    \toprule
    \toprule
    &Region & $\Gamma$ & \multicolumn{2}{c}{Norm$^a$}&${\chi}^2_{\textrm{red}}$(dof) \\
    \midrule
    \multirow{2}{*}{\textit{Chandra}} &Tail$^b$ & 2.01$^{+0.39}_{-0.38}$ & \multicolumn{2}{c}{$9.03^{+6.58}_{-3.79} \times 10^{-5}$} & 0.98 (53)\\
    &Outflow  & $1.24^{+0.23}_{-0.24}$ &\multicolumn{2}{c}{$0.80^{+0.23}_{-0.21} \times 10^{-5}$} & 1.01 (82)\\
     \midrule
    \multirow{6}{*}{\textit{XMM-Newton}}&Tail&  $1.84^{+0.19}_{-0.18}$ &\multicolumn{2}{c}{$4.36^{+1.47}_{-1.11}\times 10^{-5}$}& 1.29(118) \\
    &~~CN & $1.75.^{+0.22}_{-0.21}$ &\multicolumn{2}{c}{$2.67^{+1.06}_{-0.77}\times 10^{-5}$}& 1.13(80)\\
    &~~trail & $1.86^{+0.40}_{-0.38}$ &\multicolumn{2}{c}{$1.00^{+0.83}_{-0.46}\times 10^{-5}$}& 1.33(29) \\
    &Outflow  &  $1.37^{+0.31}_{-0.33}$ &\multicolumn{2}{c}{$0.13^{+0.06}_{-0.05}\times 10^{-5}$}& 0.75(46) \\
    &Northern halo  &  $1.68^{+0.24}_{-0.24}$ &\multicolumn{2}{c}{$4.16^{+1.92}_{-1.33}\times 10^{-5}$}& 1.91(21) \\
    &Southern halo  &  $1.44^{+0.32}_{-0.32}$ &\multicolumn{2}{c}{$8.20^{+5.15}_{-3.23}\times 10^{-5}$}& 1.34(52) \\
    \midrule
    \multirow{4}{*}{\textit{NuSTAR}}&&&Norm FPMA&Norm FPMB&\\
    &Region A& $2.04^{+0.37}_{-0.36}$ &$2.49^{+2.47}_{-1.30}\times 10^{-5}$ & $2.39^{+2.34}_{-1.23}\times 10^{-5}$&$0.40(21)$ \\
    &Region C & $2.48^{+0.31}_{-0.30}$ & $7.90^{+5.58}_{-3.34}\times 10^{-5}$& $9.16^{+7.05}_{-4.13}\times 10^{-5}$ &$1.28(36)$ \\
    &Region D & $1.90^{+0.26}_{-0.26}$ & $10.96^{+6.98}_{-4.42}\times 10^{-5}$ & $3.03^{+2.79}_{-1.62}\times 10^{-5}$ &$1.11(125)$ \\ 
    \bottomrule
    \multicolumn{6}{l}{\footnotesize \textbf{Note}: $^a$ photons/keV/cm$^2$/s @ 1 keV. $^b$ fitted from 3.0-8.0 keV to remove the thermal emission residue}
    \end{tabular}
    \label{tab:spectral}
\end{table*}

\subsection{Outflow in BSPWN B1853+01}

The presence of an outflow preceding PSR B1853+01 clearly visible in hard X-ray energy band (above 2 keV) taking a morphology of an antennae-like structure is puzzling. The verification of the nature of this emission as non-thermal ($\Gamma_{Chandra}=1.24^{+0.23}_{-0.24}$) and its association with the pulsar could have strong implications for our understanding of the system. 

Jet-like structures have been detected in PWNe associated with supersonically moving pulsars, e.g. in Geminga \citep{Posselt2017}, PWN B1929+10 \citep{Hui2008A&APSRB1929} and PWN J1153+6055 \citep{Bordas2021}. These structures could be interpreted in the context of genuine pulsar jets, which are found to be bent by the ram pressure of the surrounding medium, forming the observed arc-shaped structures. The cometary tail trailing the pulsar in these systems corresponds to a motion-distorted torus, or as part of the limb-brightened shocked pulsar wind. Recent numerical simulations of fast-moving PWNe \citep{Barkov2019b}, where the influence of geometric factors on the magnetic structure in the shocked pulsar wind are consistently addressed, provide further support to this scenario. On the other hand, misaligned X-ray outflows detected in a number of BSPWNe are instead difficult to interpret as pulsar jets or shocked pulsar wind structures, such as the filaments detected in the Guitar Nebula \citep{Hui_Becker2007Guitar, Wang2021RNAAS}, the Lighthouse Nebula \citep{Pavan2014,Pavan2016}, PWN~J2030+4415 \citep{deVries2022ApJ}, the northern outflow in PWN~J1509-5850 \citep{Klingler2016J1509} or the ``whisker'' features in the Mushroom Nebula (PWN B0355+54) \citep{Klingler2016mushroom}. These non-bent misaligned features may be produced instead by high-energy particles escaping the PWN bow shock region through reconnected magnetic field lines, emitting synchrotron emission that traces the ISM magnetic field, a scenario originally proposed in \citet{Bandiera2008}. 

Highly energetic particles might be able to escape the PWN when their gyro-radii is comparable or exceeds the stand-off distance of the bow shock. For electrons emitting synchrotron radiation at a few keV in a (possibly amplified) magnetic field $B \sim$ few~$\times 10~\mu$G (see e.g. the derived values for the Guitar and the Lighthouse nebula in \cite{2024A&AOlmi}), Lorentz factors of $\gamma \sim 10^8$ are required. These electrons will therefore be able to escape the parent PWN if the bow shock is shorter than their gyro-radii, which at these energies $\gamma$ and $B$ values is $r_g \sim$~few~$\times \, 10^{15}$~cm. Although this particle leakage mechanism is supported by recent 3D MHD simulations \citep{Barkov2019b, Olmi2019b}, our \textit{Chandra} data cannot constrain the bow shock radius further than $\sim R_{TS} \gtrsim 0.01$~pc, or about $\sim 3 \times 10^{16}$~cm, as derived above.

PSR B1853+01 is moving at the edge of the thermal X-ray emission region inside of the SNR W44, where the pulsar possibly has experienced or is experiencing a relatively abrupt change of the environment conditions. The sudden decrease in density of the surrounding ejecta might not be able to effectively confine particles in the shocked wind anymore. Another possibility is that the pulsar wind shock has already interacted with the reverse shock, leading to an enhancement of high-energy particles escaping the bow shock region and streaming farther out. In this scenario, the orientation of the outflow feature observed in PWN~B1853+01 might indicate the local magnetic field structure. The magnetic field geometry inside of W44 inferred from both CO and HCO$^+$ seems indeed oriented along the north-east direction and showing almost almost insignificant intensity variations \citep{Liu2022MNRAS}.

The spectral analysis in Sect.~\ref{spectralresults} reveals that the outflow feature in PWN B1853+01 has a relatively harder spectrum compared to that of the pulsar or the PWN tail ($\Gamma_\textrm{outflow}=1.24^{+0.23}_{-0.24} < \Gamma_\textrm{PSR}=1.87^{+0.48}_{-0.43}$ / $\Gamma_\textrm{tail}=2.01^{+0.39}_{-0.38}$ from \textit{Chandra}). In fact, misaligned outflows discovered in other BSPWNe sources usually display relatively hard spectra ($\Gamma_{\textrm{B2245}}=1.14^{+0.34}_{-0.31}$,  $\Gamma_{\textrm{J1101}}=1.7\pm0.1$, $\Gamma_{\textrm{J2030}}=1.18\pm0.20$, $\Gamma_{\textrm{J1509}}=1.50\pm0.20$, $\Gamma_{\textrm{B0355}}=1.60^{+0.32}_{-0.29}$ in the 0.5 to 8 keV band). This could be a fingerprint of the energy-selection process leading to the formation of such structures in a particle-escape scenario.
 
It is also worth noting that in the Galactic Center region, a number of X-ray filaments have been detected by \textit{Chandra} and \textit{XMM-Newton} \citep{Johnson2009MNRAS, Wang2002ApJ}. Numerous narrow radio filaments have been reported by MeerKAT \citep{Heywood2022ApJ} (see also \citet{Khabibullin2024MNRASradio}, who reported the discovery of a radio filament located ahead of PSR J0538+2817). These filaments could be streams of high-energy particles escaping yet unidentified PWNe, with the radio ones representing the low-frequency analogues of those seen in X-rays (\citealp{Barkov2019c}, \citealp {Bordas2021}).

\subsection{Halos in BSPWNe}

The faint, halo-like extended X-ray emission encompassing the central core PWN B1853+01, is particularly notable as it follows closely the morphology of the central PWN in almost all directions. Specifically, the southern part of the ``halo'' takes precisely the same shape as the outflow feature observed just ahead of the pulsar. The detection of this large X-ray halo is, in fact, consistent with the results obtained from ASCA observations, see e.g. Fig. 1 in \citet{Harrus1996}, in which an elongation of the PWN towards the south in both the 4.0--9.5 keV GIS and SIS maps is apparent. These authors then noted that the position of the hard X-ray source they detected at the position of the pulsar is marginally inconsistent with the peak position of the diffuse radio nebula to the north. The 4-8 keV MOS+\textit{pn} image of SNR W44 in \citet{Okon2020ApJ} also strongly suggests the existence of a large-scale structure near the pulsar/PWN position. However, the PWN related hard X-ray emission was not discussed there. 

The extended X-ray emission associated with PWN B1853+01, exhibits on the other hand striking similarities with that of the Snail PWN inside of SNR G327.1-1.1, the archetype of a PWN/SNR system in its late-time evolution stage. In the core nebula of the Snail PWN, a compact source is located at the apex of a tail of X-ray emission that extends in the southeast direction while two prong-like structures protrude towards northwest from the head of the bow shock. The whole core nebula is embedded in a faint diffuse emission where the two prong-like structures seem to be extending into large arcs \citep{Temim2009ApJ,Temim2015ApJ}. Note that extended TeV emission from this region has been reported (HESS J1554-550, \citealp{HESS_Snail_2018A&A}). The physical size of this diffuse TeV emission is approximately equivalent to that of the TeV halo discovered around the Geminga PWN.

A scenario in which energetic electrons having escaped the PWN are diffusing into the ambient medium could explain the faint halo emission observed in X-rays significantly larger than the size of the PWNe and, in the case of Snail PWN, the emission in the TeV band. In the 1LHAASO catalog as well as in the 2HWC catalog, a large fraction of sources are claimed to be spatially coincident with pulsars/PWNe \citep{LHAASO2024ApJS,2HAWC_CATALOG_2017ApJ}, suggesting that TeV halo or particle escape from PWNe is a generic feature related to pulsar/PWN emission. The particle escape process could possibly first appear in the late evolution stages of the PSR/SNR system (e.g. in BSPWN B1853+01 and in the Snail PWN) due to the passage of the reverse shock which crashed the quasi-steady PWN \citep{Giacinti2020A&A}. These electrons diffuse into the more distant region where they interact with the ambient photon field and are responsible for the IC radiation observed in the TeV energy band and also possible extended synchrotron radiation in X-rays whereas the ratio of the X-ray and TeV halo intensities $P_{\text{syn}}/P_{\text{IC}} = {u_B}/u_{\text{ph}}$ would depend on the ratio of the energy density between the magnetic field and the scattering photon field \citep{Aharonian1997MNRAS,Linden2017PhRvD}.

So far no confident detection of emission at GeV or TeV energies have been reported from PWN B1853+01. The 95\% confidence contour for the EGRET source 3EG 1856+0114 coincides with the southeastern sector of the SNR W44 where PWN B1853+01 is located \citep{Thompson1996ApJS}. While the reported GeV was interpreted by \citet{deJager1997ApJ} in terms of relativistic bremsstrahlung and IC scattering produced by a power-law distribution of relativistic electrons injected by PSR B1853+01, an alternative possibility is that a significant part of the gamma-ray flux observed may also be due to SNR shell-cloud interactions. In this regard, GeV emission has been recently reported from two extended structures located at the two opposite ends of the remnant along its majors axis \citep{Uchiyama2012ApJ,Peron2020ApJ}. Interestingly, the southern structure lies close to the position of the PWN, although its spectral properties and vicinity to enhanced gas density regions strongly suggest a SNR-cloud interaction for its origin (\citealp{Peron2020ApJ}). On the other hand, VHE gamma-ray fluxes at the level of 0.1 Crab are expected to be detected around PSR B1853+01 \citep{Aharonian1997MNRAS} yet no VHE gamma-ray observation pointed to the immediate vicinity of the PWN has been reported so far. VERITAS has reported upper limits to the VHE flux from the W44 region at a level $\sim$ 1.2 $\times$ 10 erg s$^{-1}$ cm$^{-2}$ in the 0.5–5.0 TeV range \citep{VERITASW442025ApJ...983...73A}. MAGIC, on the other hand, recently studied the northwestern region of W44, with no firm detection of any VHE counterpart \citep{Abe2025}. No sources coincident with W44 nor the HE structures however were identified neither in the first LHAASO catalog (\citealp{Cao2024}) nor in the third HAWC catalog (\citealp{Albert2020}).

\section{Conclusions}
\vspace{0.25cm}
\noindent We have presented a comprehensive analysis of the X-ray emission around PSR B1853+01 using data from \textit{Chandra}, \textit{XMM-Newton} and \textit{NuSTAR}. An X-ray PWN around PSR B1853+01 of rather complex structures is revealed: a cometary tail stretching along the northwest-southeast direction, of $\sim$1.5\arcmin~($\sim$1.4 pc) in size, and a small protruding outflow feature, extending toward the south $\sim$1\arcmin~($\sim$1.0 pc). \textit{XMM-Newton} observation analysis on the other hand for the first time unveiled a faint X-ray halo feature encompassing the central PWN. The southern part of the halo structure, reaching about 6\arcmin~($\sim$ 6.0 pc), morphologically resembles the outflow feature, being larger in size and fainter in brightness. At the same position of the southern part of the halo, X-ray emission is also detected with \textit{NuSTAR}. And for the first time the PWN surrounding PSR B1853+01 is detected above 10 keV up to about 20 keV.

We confirmed a non-thermal origin for the outflow feature and for the halo-like X-ray emission, displaying a harder spectrum compared to that of the pulsar or the PWN tail ($\Gamma_\textrm{outflow}=1.24^{+0.23}_{-0.24} < \Gamma_\textrm{PSR}=1.87^{+0.48}_{-0.43}$ / $\Gamma_\textrm{tail}=2.01^{+0.39}_{-0.38}$ from \textit{Chandra}, $\Gamma_\textrm{halo\_S}= 1.44^{+0.32}_{-0.32}<\Gamma_\textrm{tail}=1.84^{+0.19}_{-0.18}$ from \textit{XMM-Newton}). We thus propose that both the outflow feature and the halo-like emission region surrounding the whole PWN nebula are produced by energetic particles escaping the PWN bow shock region.

X-ray instruments of next generation, such as \textit{AXIS} \citep{AXIS2023}, with excellent angular resolution across a wide FoV and improved flux sensitivity would help to constrain the size of the X-halo structure and, eventually, closely pinpoint the sites of particle acceleration and escape. Finally, radio interferometric observations of BSPWN B1853+01 might help to better constrain the pulsar proper motion velocity and the location of the standoff distance, further constraining the spectrum of the escaping particles.

\vspace{0.25cm}
\begin{acknowledgements} 
XZ acknowledges funding by l'Agència de Gestió d'Ajuts Universitaris i de Recerca (AGAUR) under the FI-SDUR grant (2021 FISDU 00369). PB and XZ also acknowledges support from Ministerio de Ciencia, Innovación y Universidades (MCIUN) through Proyectos I+D Generación de Conocimiento PID2019-105510GB-C31, PID2022-136828NB-C41 and PID2022-138172NB-C43.
SH's research is primarily supported by the Natural Sciences and Engineering Research Council of Canada (NSERC) through the Discovery Grants and Canada Research Grants Programs. KI acknowledges support under the grant PID2022-136828NB-C44 provided by MCIN/AEI/10.13039/501100011033/FEDER, UE. Authors also thank George Pavlov for his valuable comments that helped to improve the paper.\end{acknowledgements}

\bibliographystyle{aa}
\bibliography{w44pwn.bib}
\end{document}